\documentstyle[twoside,fleqn,espcrc2]{article}

\def\be{\begin{equation}}
\def\ee{\end{equation}}
\def\bea{\begin{eqnarray}}
\def\eea{\end{eqnarray}}
\def\lsim{\mathrel{\rlap{\lower4pt\hbox{\hskip1pt$\sim$}}
    \raise1pt\hbox{$<$}}}         
\def\gsim{\mathrel{\rlap{\lower4pt\hbox{\hskip1pt$\sim$}}
    \raise1pt\hbox{$>$}}}         
\newcommand{\dst}{\displaystyle}

\newcommand{\fr}[2]{\frac{{\dst #1}}{{\dst #2}}}
\newcommand{\epe}{\mbox{$e^+e^-\,$}}
\newcommand{\ggam}{\mbox{$\gamma\gamma\,$}}
\newcommand{\egam}{\mbox{$\gamma e\,$}}
\newcommand{\eeww}{\mbox{$e^+e^-\to W^+W^-\,$}}
\newcommand{\ggww}{\mbox{$\gamma\gamma\to W^+W^-\,$}}

\newcommand{\egeh}{\mbox{$e\gamma\to eH\,$}}
\newcommand{\gewnu}{\mbox{$e\gamma\to W\nu\,$}}
\newcommand{\SM}{${\cal S} {\cal M} \;$}
\newcommand{\DSM}{$2{\cal H} {\cal D} {\cal M} \;$}
\newcommand{\MSM}{${\cal M} {\cal S}  {\cal M}\;$}
\newcommand{\CP}{${\cal C}{\cal P}\;$}
\newcommand{\MSSM}{${\cal M}{\cal S} {\cal S}  {\cal M}\;$}
\newcommand{\SU}{${\cal S}{\cal U} 2\otimes{\cal U} 1\;$}

\newenvironment{Itemize}{\begin{list}{$\bullet$}%
{\setlength{\topsep}{0.2mm}\setlength{\partopsep}{0.2mm}%
\setlength{\itemsep}{0.2mm}\setlength{\parsep}{0.2mm}}}%
{\end{list}}
\newcounter{enumct}
\newenvironment{Enumerate}{\begin{list}{\arabic{enumct}.}%
{\usecounter{enumct}\setlength{\topsep}{0.2mm}%
\setlength{\partopsep}{0.2mm}\setlength{\itemsep}{0.2mm}%
\setlength{\parsep}{0.2mm}}}{\end{list}}

\newcommand{\AmS}{{\protect\the\textfont2
  A\kern-.1667em\lower.5ex\hbox{M}\kern-.125emS}}
\title{Physics at \ggam and \egam colliders}

\author{Ilya F. Ginzburg\address{Sobolev Institute of Mathematics,
Siberian Branch of RAS,\\
Prosp. ac. Koptyug, 4, 630090 Novosibirsk, Russia}}%

\begin{document}
\begin{abstract}
I discuss, what really new could give Photon Colliders
($\gamma\gamma$ and $e\gamma$) after LHC and \epe Linear Collider
operations.
\end{abstract}

\maketitle

\section{Possible scenarios}

$\bullet$ The developed physical programs of new machines are
based on the idea, that {\large\em the Nature is so favorable to
us that it has placed the essential fraction of new particles
within the LHC operation domain.} Moreover, in these programs we
believe, that new particles or interactions belong to one of the
known sets and we consider an opportunity to see them. Together
with Higgs boson (or bosons) in the one or two doublet \SM, the
expected most probable variants are SUSY--\MSSM, leptoquarks,
$WW$ or $WZ$ resonances, compositeness, effects from large extra
dimensions,...  Such type assumption is necessary component of
the physical program of the LHC, where the discovery of
something unexpected is hardly probable.

Main goal of the LHC is a discovery of some of enumerated
particles or effects. The \epe Linear Colliders should measure
parameters of new particles and corresponding couplings with
high accuracy. Some unexpected new particles will be seen here
via the threshold behaviour in two jet or lepton production.
The Photon Colliders in this approach are considered as the
machines for the precise study of QCD and cross check of
couplings measured earlier.

$\bullet$ In this approach one forget some problems of
the \SM (including its foundation).  The solutions (still
unknown) could be tested with the aid of Photon Colliders only.

$\bullet$ We should be ready to meet the opposite opportunity:
{\large\em no new particles will be discovered at LHC, except
Higgs boson} (\SM or \DSM or \MSSM). In this case Photon
Colliders could give the best key to the discovery of a New
Physics.

$\bullet$ In this report we consider this very opportunity. We
assume that either New Physics differs strong from the expected
variants or new particles in the one of expected variant are very
heavy (except Higgs boson(s)) or some difficult opportunity within
\MSSM is realized (see e.g. \cite{Kras}).

\section{Photon Colliders. Main features}

The future Linear Colliders would form the complexes including both
\epe collider mode and Photon collider mode (\ggam and \egam)
\cite{SLACDESY,GKST}. This photon mode is based on the \epe\  one
with electron energy $E$ and luminosity ${\cal L}_{ee}$.  Here I
present its main characteristics {\em assuming no special efforts in
optimization of photon mode at the initial stages of acceleration}
\cite{SLACDESY,GKST}.

\begin{Itemize}
\item Characteristic photon energy $E_\gamma\approx 0.8E$.
\item Annual luminosity  ${\cal L}_{\gamma\gamma}\approx 100$
fb$^{-1}$  (${\cal L}_{\gamma\gamma}\approx 0.2{\cal L}_{ee}$).
\item Mean energy spread $<\Delta E_\gamma>\approx 0.07E_\gamma$.
\item Mean photon helicity $<\lambda> \approx 0.95$ with easily
variable sign. One can transform this polarization into the linear
one \cite{Kotser}.
\item There are no special reasons against observations at small
angles except non--principal technical details of design.
\item The conversion region is \egam collider with c.m.s. energy
about 1.2 MeV but with annual luminosity $\sim 10^6$ fb$^{-1}$!
\end{Itemize}
Below we denote by $\lambda_i$ mean degree of the photon circular
polarization (helicity) and by $\ell_i$ mean degree of their linear
polarization, $\phi$ is the angle between directions of these linear
polarizations for opposite photons, $y_i=E_{\gamma i}/E$.

{\large\bf Approximate photon spectra \cite{Ginkot}.} The luminosity
distribution in the effective $\gamma\gamma$ mass has usually two
well separated peaks:
 \newline
{\em a) high energy peak} with mentioned mean energy spread $7\%$
and integrated luminosity $0.2{\cal L}_{ee}$;
 \newline
{\em b) wide low energy peak.} The last depends strongly on
details of design, it is unsuitable for the study of New Physics
phenomena.

The form of high energy peak depends on the reduced distance
between conversion and interaction points $\rho$. At $\rho=0$
separation of peaks is practically absent. In the modern projects
$\rho\approx 1$. In this case peaks are separated well, the high
energy peak contains about 30 \% of geometrical luminosity and 80
\% from luminosity of peak at $\rho=0$. At $\rho^2<1.3$ and the
ellipticity of beam $A>1.5$ the high energy peak is independent
on $A$ (usually $A\gg 1$). In this region the form of high
energy peak is approximated with high precision by a convolution
of two effective photon spectra with functions
$\tilde{F}(y,\rho)$ written in ref.~\cite{Ginkot}
\be
d{\cal L}=\tilde{F}(y_1,\rho)\tilde{F}(y_2,\rho)dy_1dy_2\,.
\ee

\section{Higgs window to a New Physics}

The basic point here is the opportunity to measure the two--photon
width of the \MSM Higgs boson with accuracy 2 \% \cite{JikS} (one can
hope to improve this accuracy since the used luminosity integral 30
fb$^{-1}$ corresponds to only three month operations).

$\bullet$ Let us first talk about the opportunity that some of the
discussed scenarios (\DSM, \MSSM) is realized, but new particles are
heavier than those observable at the LHC (decoupling regime) with two
variants:
\begin{Enumerate}
\item {\em Strong decoupling}. The additional Higgs bosons
$H,\;A,\;H^\pm$ are also very heavy.
\item {\em Weak decoupling}. The additional Higgs bosons
$H,\;A,\;H^\pm$ are lighter than 400 GeV.
\end{Enumerate}

We distinguish the following variants of \SM or its extensions:
\newline
$\star$ \SM with one Higgs boson doublet --- \MSM;
\newline
$\star$ \SM with two Higgs boson doublet --- \DSM, or its SUSY
extension --- \MSSM in the decoupling regime.
\newline
In this regime the difference between the \MSSM and the \DSM in
the \MSSM like variant (at $M_{H^\pm}^2\approx \lambda_5v^2$) is
hardly observable. So we denote usually both cases as \MSSM. We
denote by \DSM only its general variant with $M_{H^\pm}^2\gg
\lambda_5v^2$. Our discussion below is based on the results of
Refs.~\cite{Djou,GKO}.

The studies at the LHC could give us some coupling constants of the
Higgs boson with a matter (quarks, leptons, $W$ and $Z$) with
accuracy about 10~\%. The measurements at \epe Linear Collider will
improve accuracy to the level of about 1~\%.  The deviation of these
couplings from their \MSM values could be considered as a signal for
realization of a \DSM or \MSSM. The measurement of the two--photon
width of a (lightest) Higgs boson allows to separate the general
\DSM from the \MSSM. The additional measurement of a $HZ\gamma$
coupling in the \egeh process \cite{Il,BGI} could support this
differentiation.

We show this for the difficult enough case.  Let the measured
couplings of the Higgs boson with a matter are given by the
\MSM. The same values can be obtained in the \DSM or
\MSSM at $\beta-\alpha=\pi/2$. In the $H\ggam$, $HZ\gamma$ widths
effects of $W$ and $t$ quark loops are of opposite sign. That is why
the effect of very heavy charged Higgses is enhanced here up to about
13\% in the general \DSM. This difference will be seen good even in
the case of strong decoupling. In the \MSSM this effect reduces to a
few percent level (just as the effect of heavy superparticles) ---
depending of the mass of these particles.

Therefore the measurement of the Higgs boson \ggam width with 2 \%
accuracy could answer which model is realized --- general \DSM or
\MSSM. And one can get this answer before the discovery of
superpartneurs. This measurement could differentiate general \DSM
from \MSM together with \MSSM at $\beta-\alpha=\pi/2$. The
discrimination of the \MSSM from the \MSSM like \DSM needs higher
precision in the measuring of the discussed width.

In the weak decoupling case the additional measurements of
$H\ggam$, $A\ggam$ couplings in the \ggam\ collisions,
$HZ\gamma$, $AZ\gamma$, $H^-W^-\gamma$ couplings in the \egam\
collisions would be very useful to discriminate the discussed
opportunities.

$\bullet$ Let the New Physics is different from the discussed
models. When the collision energy is below scale of New Physics
$\Lambda$, last manifests itself via {\em anomalies} in the
interactions of known particles. In this case {\bf Photon
Colliders provide the best place for discovery of this New
Physics and for understanding it.} Indeed the $H\ggam$, $H\gamma
Z$,...  vertices in the \SM are one--loop effects. Therefore the
relative value of anomalies is enhanced here in comparison with
other interactions.

Together with the \SM effects the above anomalies are described by an
Effective Lagrangian relevant photon collisions:
\be\begin{array}{c}
{\cal L}_{H\gamma}=\fr{ G_\gamma
HF^{\mu\nu}F_{\mu\nu}}{2v} +\fr{G_Z H F^{\mu\nu}Z_{\mu\nu}}{v}
+\\
\fr{ \tilde{G}_\gamma
HF^{\mu\nu}\tilde{F}_{\mu\nu}}{2v} +\fr{\tilde{G}_Z H
\tilde{F}^{\mu\nu}Z_{\mu\nu}}{v}\,.\\
G_i=\fr{\alpha\Phi_i^{SM}}{4\pi}+\theta_i\fr{v^2}{\Lambda_i^2}\,,
\quad (i=\gamma\,,\;Z)\,.
\end{array}\label{higano}
\ee
Here $Z_{\mu\nu}$, $F_{\mu\nu}$ and $\tilde{F}_{\mu\nu}$ are standard
field strength tensors, $v=246$ GeV.  The values $\Phi_i^{SM}$ are
well known $(|\Phi_i|\sim 1)$. For the \CP even case $\tilde{G}_i=0$,
$\theta_i =\pm 1$.  For the \CP odd case all quantities $\theta_i$,
$\tilde{\theta}_i$ can be complex, $\theta_a=e^{\dst i\phi_a}$.

The \CP odd anomalies manifest itself in the polarization asymmetry
in the production $\ggam\to H$, $\egeh$. In particular, for the
$\ggam\to H$ process we have \cite{CPGI}
\be\begin{array}{c}
<\sigma>(\lambda_i,\ell_i,\psi)=<\sigma^{SM}>_{np}
\fr{T(\lambda_i,\ell_i\psi)}{|G_\gamma^{SM}|^2};\\
T(\lambda_i,\ell_i,\psi)=|G_\gamma|^2 (1+\lambda_1\lambda_2
+\ell_1\ell_2 \cos 2\psi)\\
  +|\tilde{G}_\gamma|^2
\left(1+\lambda_1\lambda_2-\ell_1\ell_2 \cos 2\psi \right)\\ +
2Re(G_\gamma^*\tilde{G}_\gamma)(\lambda_1+\lambda_2) + 2
Im(G_\gamma^*\tilde{G}_\gamma)\ell_1\ell_2\sin 2\psi\,.
\end{array}\ee

Similar equations were obtained for the \egeh process and $HZ\gamma$
anomalies. The sensitivity of the corresponding experiments to the
scale of New Physics was studied in Refs.~[9--13].

\section{Gauge boson physics}

$\bullet$ In the discussed energy range the New Physics effects
will be seen as some deviations from the prediction of \SM.
These deviations can be described by anomalies in the Effective
Lagrangian ${\cal L}_{eff}$. There are many anomalies of
dimensions 6 and 8 relevant to the gauge boson interactions (\CP
even and \CP odd). Each process reacts for several of them.  The
separate extraction of different anomalies is difficult for the
\epe mode with only one well measurable process \eeww. (The
$\gamma WW$ and $ZWW$ vertexes enter this process
simultaneously.)

In these problems the potential of the Photon Collider is
exceptional. Indeed, large variety of the processes with the
production of gauge boson will be observed here with both high
purity and counting rate (millions or hundred thousands events per
year) (\ggww, \gewnu, $\gamma e\to eWW$, $\egam\to \nu WZ$,
$\ggam\to WWZ$, $\ggam\to WWWW$, $\ggam\to WWZZ$, $\egam\to\nu
e^+e^-W$, ...). This large variety allows to separate different
anomalies. For example, one can realize such a program: First, to
extract $\gamma WW$ anomalies from $\egam\to \nu W$ process. After
that, to extract $ZWW$ anomalies from \eeww process. Last, to
extract $\ggam WW$ anomaly from \ggww. In the same way the process
$\egam\to eWW$ allows to study anomaly $\gamma ZWW$.

The study of some separate distributions in the specific regions
of parameter space can enhance effect of some anomalies (in
comparison with the entire cross section). For example, the
total cross section of the process $\egam\to eWW$ is determined
mainly by effect of \ggww subprocess, the $Z\gamma\to WW$
subprocess become very essential in the cross section of this
process, given at the transverse momentum of an electron
$p_\bot>30$ GeV; the polarization asymmetry is sensitive to the
possible \CP odd anomalies.

$\bullet$ The dynamical \SU symmetry breaking is also considered
often. Here this breaking is caused the strong interaction of
$W$ bosons (longitudinal components) --- instead of Higgs bosons
at $E\gsim 4\pi v\approx 3$ TeV. At lower energies amplitudes
differ weakly from their \SM values.

The \egam collider allows study this strong interaction at
smaller energy. For this goal, one should study the process
$\egam\to eWW$ and should consider charge asymmetry of produced
$W$'s (longitudinally polarized, if possible), caused
interference between t--channel $\gamma/Z$ exchange diagram and
corresponding bremsstrahlung diagram.  Even if the cross section
itself differs weakly from its \SM value, the value of this
asymmetry is $\propto\cos(\delta_0-\delta_1)$ where $\delta_i$
are the phases of strongly interacted $WW$ amplitudes
\cite{GinSD}.

 $\bullet$ The~reactions~\ggww~and~\gewnu
\newline
 will give about 10
millions $W$'s per year. It provides an opportunity to measure
 the corresponding cross sections with two loop accuracy.

The EW theory is the standard QFT based on the whole set of the
asymptotical states for the fundamental particles. It is the
basis for the construction of perturbation theory with the
standard particle propagators.  But the fundamental particles of
theory ($W$, $Z$, $H$) are unstable.  The QFT with unstable
fundamental particles is unknown till now.  Without such a
theory precise description of EW processes is impossible.

From this point of view, the breaking of gauge invariance in the
calculations of the processes with gauge boson production (like
$\eeww\to \mu\bar{\nu}q\bar{q}$) is not the main effect but the
signal on the unsatisfactory state with EW theory. This signal should
be used in the construction of satisfactory scheme.

We hope, that new features of such scheme (as compare with
constructed recipes like \cite{Olden}) will be seen at the
expected two--loop accuracy level.

The solution of this problem will be an essential step in the
construction of QFT relevant for the description of real world.

$\bullet$ The additional interesting field here is the study of
the QCD radiative corrections to the \ggww process in the Pomeron
regime (two gluon, etc. exchange between quarks from $W$ decays or
in their polarization operator at $S\gg M_W^2$).\\

 {\bf That is a large area for new work here.}

\section{The discovery of new unexpected particles}

The experiments at the LHC could discover many expected
particles but the discovery here of some unexpected particle is
very difficult task.  Assuming that the decay products of some
unexpected charged particle contain known particles, these new
particles can be discovered at Linear Collider in both \epe\ and
\ggam\ mode. The higher value of the production cross section in
\ggam\ mode comparing with the \epe\  mode compensates difference
in the luminosities in these modes. The values of these cross
sections depend on $s/M^2$, charge and spin of produced
particle. In the \epe\  mode the
additional dependence on the coupling with $Z$ make difficult an
unambiguous restoration of the charge and spin of produced
particle from the data. The \ggam\  mode is free from this
difficulty. Additionally, the polarization dependence is useful
to determine spin of produced particle independent on its
charge. Near the threshold this cross section in the \ggam mode
is
\be
\propto (1 +\lambda_1\lambda_2 \pm\ell_1\ell_2 \cos2\phi)
\ee
with sign + for scalars and sign -- for spinors.

These cross sections decreases slowly with energy growth. They are
high at large enough energies. It allows to study decay products in
the region where they don't mix with each other.

\section{ $\ggam\to \ggam$ process for the nonstandard New Physics.}

Some authors discuss the cases when this process become
observable due to loops with some new particles $F$. It happens
possible if the c.m.s. energy is larger than $2M_F$ (usually
much larger). These effects cannot give us new information about
the particles $F$ because of the processes like $\ggam\to
F\bar{F}$ have higher cross sections and they observable usually
a at lower energy.

So I discuss here only two topics related to the nonstandard New
Physics:

$\star$ Heavy point--like Dirac monopole.

$\star$ Effect of extra dimensions.

In both cases we consider the process much below new particle
production threshold. Denoting corresponding mass by $M$, the
cross section can be written in the form
\be
\sigma(\ggam\to\ggam) =\fr{A}{32 \pi s} \left(\fr{s}{4
M^2}\right)^4 \label{gggg}
\ee
with specific angular distribution (roughly --- isotropic) and
polarization dependence.  The wide angle elastic light to light
scattering has excellent signature and small QED background.

The observation of strong elastic \ggam scattering, quickly raising
with energy, will be a signal for one of mentioned mechanisms. The
study of polarization and angular dependence at photon collider can
discriminate which mechanism is working.

$\bullet$ {\bf Point--like Dirac monopole} \cite{GSh}. The
  existance of this particle explains mysterious quantization of
an electric charge. There is no place for it in modern theories
of our world but there are no precise reasons against its
existence.

At $s\gg M^2$ the electrodynamics of monopoles is expected to be
similar to the standard QED with effective perturbation
parameter $g\sqrt{s}/(4\pi M)$ where the coupling constant
$g=n/(2e)$. The effect is described by monopole loop. So, the
coefficient $A$ in eq.~(\ref{gggg}) is calculated within QED. It
is $\propto g^8$ and depends strongly on the spin of monopole
$J$ (just as details of angular and polarization dependence).
For example, $A(J=1)/A(J=0)
\approx 1900$.

The effect can be seen at TESLA500 at the monopole mass $M<4-10$ TeV
(depending on monopole spin). Modern limitation obtained at Tevatron
is about one order of value lower.

{$\bullet$} {\bf Effect of extra dimensions} \cite{dim}. Nowadays
the scenario is considered where gravity propagates in the
$(4+n)$--dimensional bulk of spacetime, while gauge and matter
fields are confined to the (3+1)--dimensional world volume of a
brane configuration. The extra $n$ dimensions are compactified
with some space scale $R$ that result Kaluza--Klein excitations having
masses $\pi n/R$. The corresponding scale $M$ in our world is
assumed to be $\sim$ few TeV. The particles of our world interact
via the set of these Kaluza-Klein excitations having spin 2 or 0.
In this approach all unknown coefficients are accumulated in the
definition of $M$ in the equation for the cross section (with
$A\sim 1$). The angular and polarization dependence of cross
section are also known.

Similar results were obtained for other processes $B{\bar B}\to
 C{\bar C}$. The two photon final state has the best signature and the
lowest \SM background. The two photon initial state has numerical
advantage as compare with \epe one.

The limitation for effect of extra dimensions correspond for the
Photon Collider based on TESLA500 is $M\approx 2.2$ TeV. It is
higher than that observable at the LHC.

\section{Axions, etc, ... from the conversion region}

Some very light and elusive particles $a$ (axions, majorons,...) are
expected in many schemes.

They can be produced in the conversion region, that is the \egam
collider with $\sqrt{s_{e\gamma_0}}\approx 1.2$ MeV or
$\sqrt{s_{\gamma\gamma_0}}\approx 1$ MeV and with annual luminosity
about million fb$^{-1}$ \cite{Pol}.

The production processes are
\be
e\gamma_0\to ea\,,\quad \gamma\gamma_0\to a\,.
\ee
The angular spread of these $a$ is very narrow and they interact with
the matter very weakly.

So, the registration scheme can be of the following type. After
the damping of photon beam one should set a lead cylinder of
diameter about 3--5 cm and length 300 m -- 1 km (within angular
spread of produced particles). After that some set of
scintillators situated in covered circle with radius about 3 m
will fix the most part of products of interaction of this
particle with the lead nuclei within the cylinder.

\section{Final notes}

{\bf The schedule in the operations of different modes and energies
of a Linear Collider depends on the results of LHC studies.}
Two variants should be considered:

$\bullet$ Let some new particles (SUSY like) will be discovered at
the LHC. In this case the natural continuation of the LC500 program
will be LC800 with Photon Colliders after that.

$\bullet$ Let no new particles (except Higgs boson) will be
discovered at the LHC. In this case the Photon Collider should
operate as soon as possible. For example, the Photon Collider with
c.m.s. energy $\sqrt{s}\approx M_h\sim 100\div 200$ GeV for the study
mainly of Higgs boson can be the first stage of entire LC project.
The advantages of this way are:

{\em The basic electron energy is lower.

The positron beam is unnecessary.}

 \section*{Acknowledgment}

I am grateful to J.J. van der Bij and S. S\"oldner--Rembold for the
kind invitation to Freiburg and this conference. This work was also
supported by grant RFBR 99--02--17211 and grant of Sankt--Petersburg
Center of Fundamental Sciences.


\begin{thebibliography}{99}
\bibitem{Kras} N.V. Krasnikov. hep--ph/9901398.

\bibitem{SLACDESY} Zeroth-order Design Report for the NLC, SLAC
Report 474 (1996); TESLA, SBLC Conceptual Design Report, DESY
97-048, ECFA-97-182 (1997); R.Brinkmann et. al.,
 NIMR {\bf A406} (1998) 13.

\bibitem{GKST} I.F.~Ginzburg,~G.L.~Kotkin,~V.G.~Serbo, V.I.~Telnov.
Pis'ma ZhETF {\bf 34} (1981) 514; Nucl.Instr.Methods (NIM) {\bf 205}
(1983) 47.  I.F.Ginzburg, G.L.Kotkin, S.L.Panfil, V.G.Serbo,
V.I.Telnov.  NIM {\bf 219} (1984) 5.
\bibitem{Kotser} G.L. Kotkin, V.G. Serbo, Phys. Lett.  {\bf
B413} (1997) 122.
\bibitem{Ginkot} I.F.~Ginzburg, G.L.~Kotkin, hep-ph/9905462.

\bibitem{JikS} M. Melles, W.J. Stirling, V.A. Khoze. hep--ph/9907238;
G.~Jikia, S.~S\"oldner-Rembold, these proceesdings.

\bibitem{Djou} A.Djouadi, V. Driesen, W. Hollik, J.I. Illana.
Eur. Phys. J. {\bf C1} (1998) 148, 163.
\bibitem{GKO} I.F. Ginzburg, M. Krawczyk, P.Olsen, in preparation.
\bibitem{Il} E.~Gabrielli, V.A.~Ilyin, B~Mele. Phys. Rev. {\bf
D56} (1997) 5945.
\bibitem{BGI} A.T. Banin, I.F. Ginzburg, I.P. Ivanov. Phys. Rev. {\bf
D59} (1999) 115001.

\bibitem{Goun} G.J. Gounaris, F.M. Renard. Z. Phys. {\bf C69} (1996) 513.
\bibitem{CPIl} E.~Gabrielli, V.A.~Ilyin, B~Mele, hep--ph/9902362.
\bibitem{CPGI} I.F. Ginzburg, I.P. Ivanov, in preparation.
\bibitem{GinSD} I.F. Ginzburg. Proc. 9th Int. Workshop
on Photon -- Photon Collisions, San Diego (1992) 474--501, World Sc.
Singapore.
\bibitem{Olden} R.G.Stuart. Phys. Lett. {\bf B267} (1991) 240, {\bf
B272} (1991) 353, Phys. Rev. Lett. {\bf 70} (1993) 3193; A.~Aeppli,
G.J.~van~Oldenborgh, D.~Wyler. Nucl. Phys. {\bf B248} (1994) 126;
H.~Veltman. Z.~Phys. {\bf C62} (1994) 35; L.~Maiani, M.~Testa. Annals
Phys. {\bf 263} (1998) 353; J.~Papavassiliou. hep--ph/9905328.


\bibitem{GSh} I.F. Ginzburg, S.L. Panfil. Sov. J. Nucl. Phys.
{\bf 36} (1982) 850; I.F.~Ginzburg, A.~Schiller. Phys. Rev. {\bf D57}
(1998) R6599; {\bf D }  (1999) in print, hep-ph/9903314.
\bibitem{dim} H.~Davoudiasi. hep--ph/9904425; K.~Cheung hep--ph/9904266.
\bibitem{Pol} S.I. Polityko, hep--ph/9905451.


\end{thebibliography}
\end{document}